\documentclass[11pt,twoside]{article}

\usepackage{asp2006}
\usepackage{epsf}
\usepackage{psfig}
\usepackage{graphicx}
\usepackage{lscape}

\begin{document}
\title{Converting H$\alpha$ Luminosities into Star Formation Rates}   
\author{Jan Pflamm-Altenburg,$^{1,2}$, Carsten Weidner,$^3$ and Pavel Kroupa$^{1,2}$}   
\affil{$^1$ Argelander-Institut f\"ur Astronommie (AIfA), Auf dem H\"ugel
  71, 
  D-53121
  Bonn, Germany}
\affil{$^2$ Rhine Stellar Dynamics Network (RSDN)}    
\affil{$^3$ Departamento de Astronom{\'i}a y Astrof{\'i}sica, 
  Pontificia Universidad Cat{\'o}lica de Chile, Campus San Joaquin, Vicu\~na
  Mackenna 4860, 782-0436 Macul, Santiago, Chile}    

\begin{abstract} 
The recent finding that the IGIMF (integrated galaxial initial 
stellar mass function) composed of  all newly formed stars in all
young star clusters has, in dependence of the SFR, a steeper slope  
in the high mass regime
than the underlying canonical IMF of each star cluster 
offers new insights into the galactic star formation process: 
The classical linear relation 
between the SFR and the produced H$\alpha$ luminosity is broken and
SFRs are always underestimated. Our new relation is likely 
to lead to a revision of the cosmological SFH.
\end{abstract}


\section{SFR-H$\alpha$-relation}
\begin{figure}
  \includegraphics[width=0.5\textwidth]{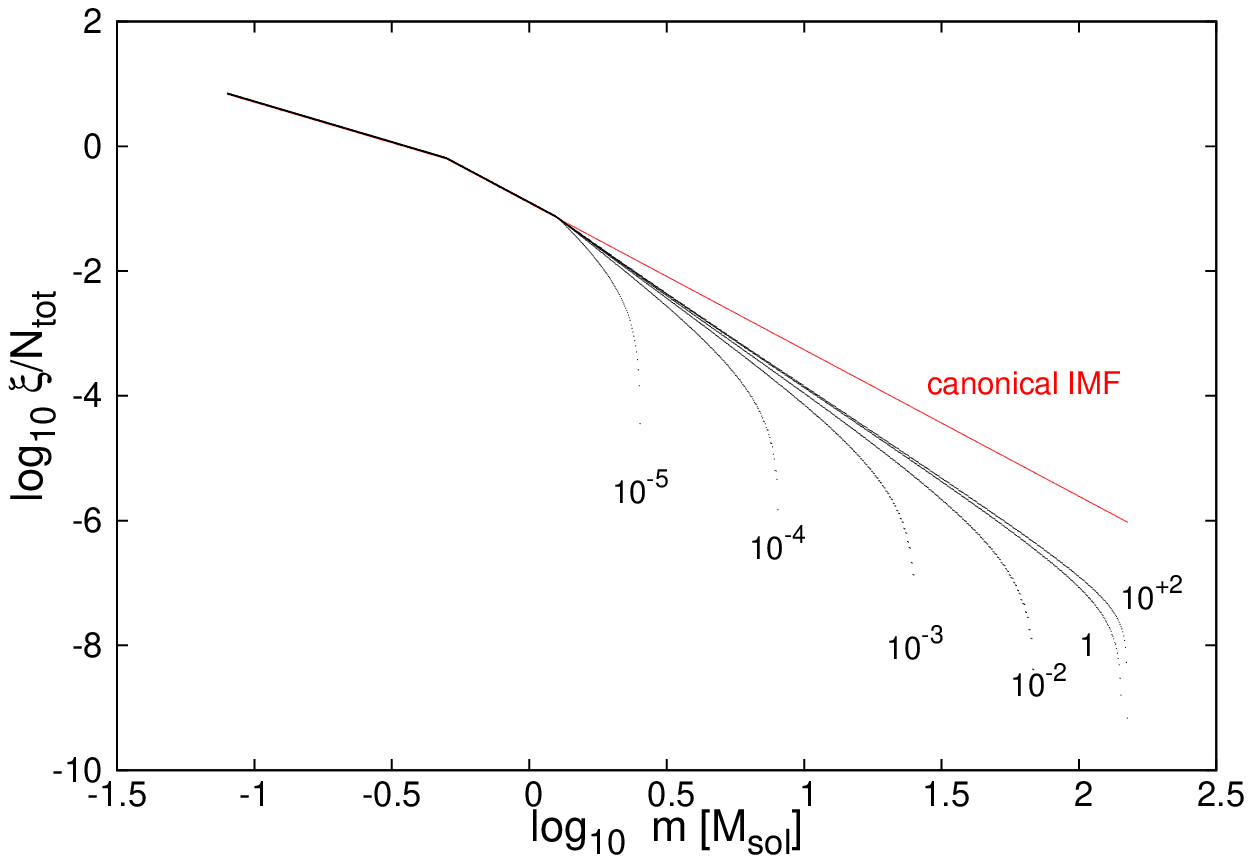}
  \includegraphics[width=0.5\textwidth]{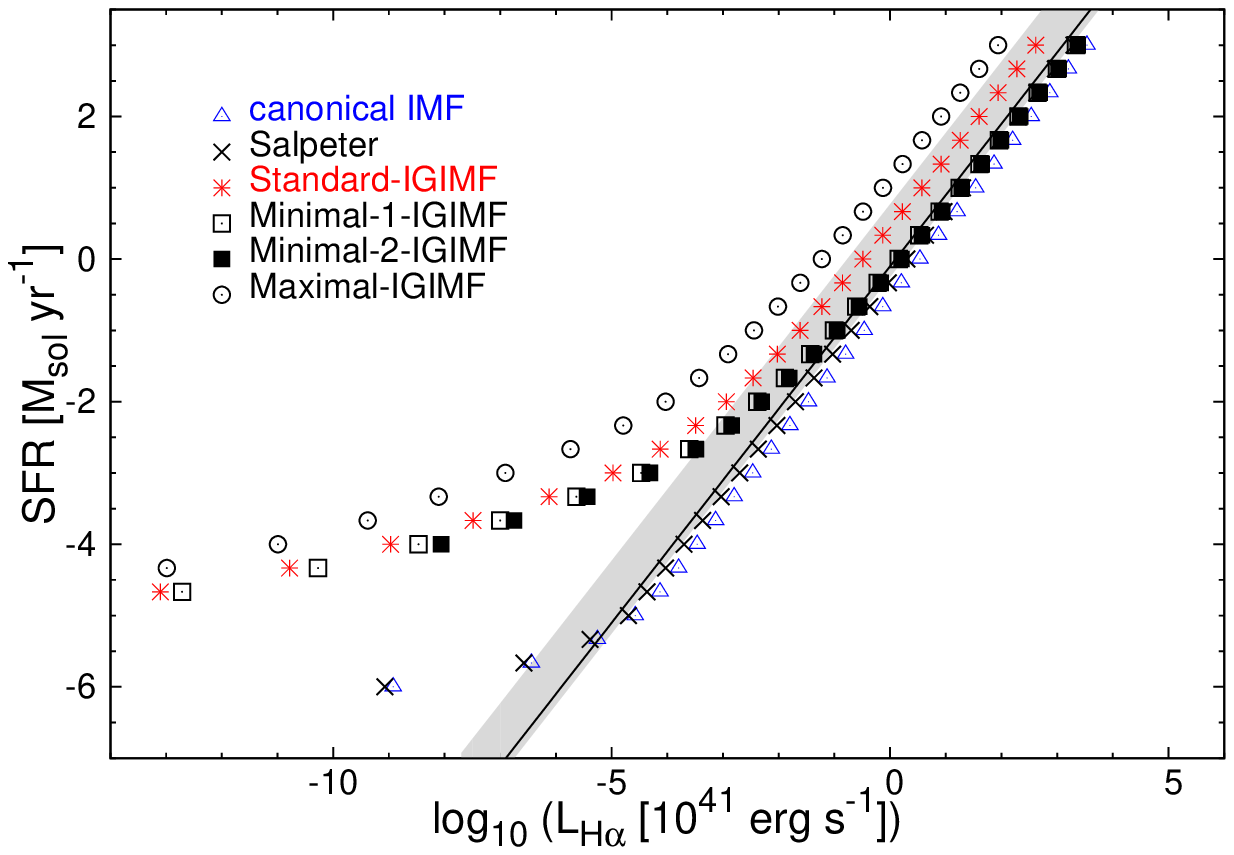}
  \caption{{\it Left}: The IGIMF in dependence of the total SFR in the
  standard scenario.
  {\it Right}: The SFR-H$\alpha$-luminosity relation for four different
  IGIMF scenarios and relations by \citet[gray shaded area]{kennicutt1994a}  
  and the widely used relation 
  $\mathrm{SFR}\;/\;\mathrm{M}_\odot\;\mathrm{yr}^{-1} = 
  L_{\mathrm{H}\alpha}\;/\;
  1.26\cdot 10^{41}\;\mathrm{erg}\;\mathrm{s}^{-1}$
  (solid line).}
  \label{fig_igimf_sfr}
\end{figure}

To obtain a relation between the total galactic wide SFR and the produced 
H$\alpha$ luminosity the IGIMF as a function of the SFR is combined 
with stellar evolution models. The resulting relations for 
four different IGIMF scenarios, introduced by  \citet{weidner2005a}, 
are plotted in Fig.~\ref{fig_igimf_sfr} (right). 
For a given H$\alpha$ luminosity the SFRs are always higher than in the
classical linear relations by \citet*{kennicutt1994a} (grey shaded area,
solid line).
In addition, two classical models are included where the IGIMF is
identical to a Salpeter IMF and a canonical IMF but using the
same stellar evolution models as used in the IGIMF models.
\section{dIrr galaxies}
\begin{figure}
  \includegraphics[width=0.5\textwidth]{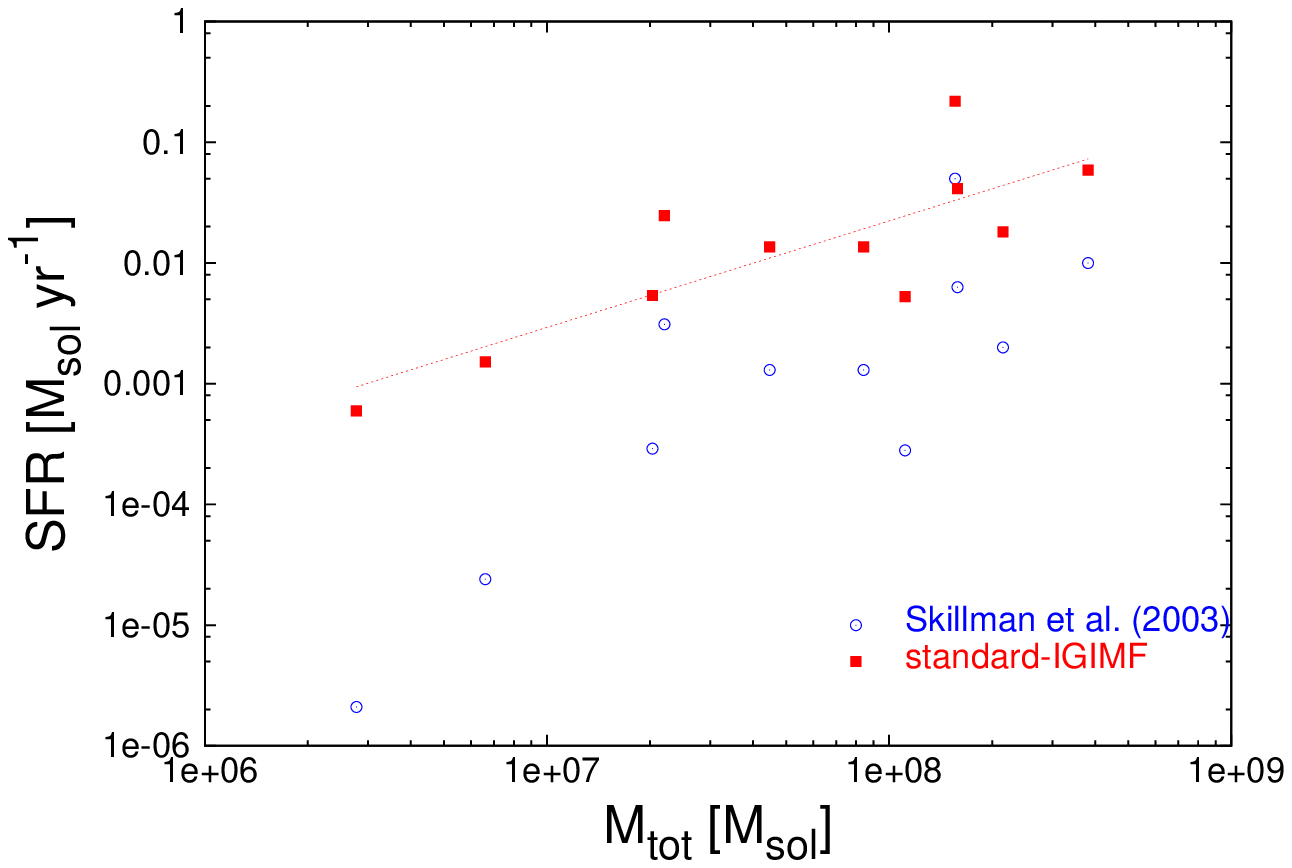}
  \includegraphics[width=0.5\textwidth]{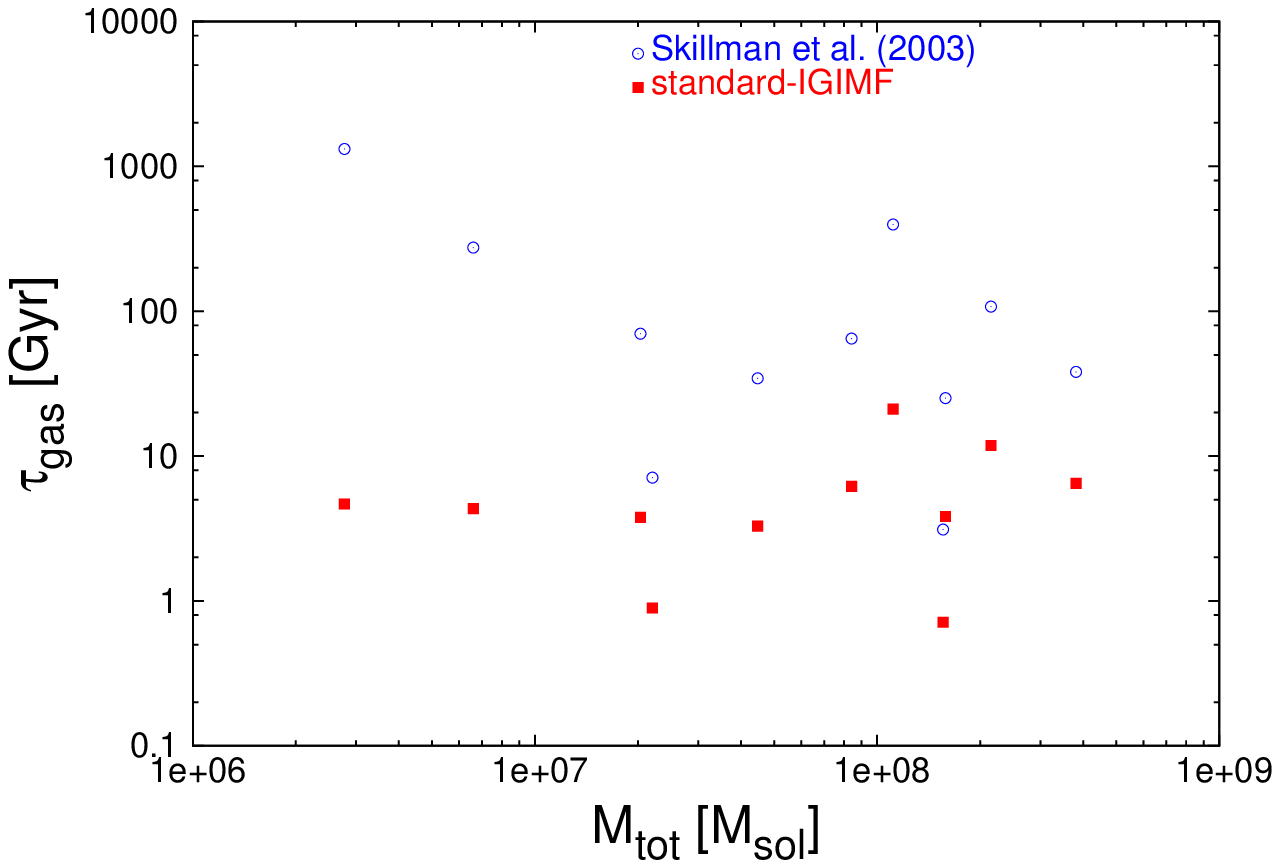}
  \caption{Derived SFRs ({\it left}) and gas depletion times ({\it right})
    of the Sculptor dIrrs based on the standard IGIMF scenario and on
    the linear Kennicutt-relation (Fig~\ref{fig_igimf_sfr}, solid line)
    (Skillman et al. 2003). Note that the standard-scenario IGIMF implies
    significantly higher SFRs and a constant gas-depletion time scale
    for all dwarf galaxies.
  }
  \label{fig_dIrrs}
\end{figure}

Applying our SFR-L$_{\mathrm{H}\alpha}$ relation to the observed
H$\alpha$-luminosities of the Sculptor dwarf irregular galaxies
\citep{skillman2003a} the SFRs (Fig.~\ref{fig_dIrrs}, left) and 
related parameters such as the gas
depletion time scale (Fig.~\ref{fig_dIrrs}, right) change dramatically.


\bibliographystyle{mn2e}
\bibliography{star-formation}


\end{document}